\documentclass[conference, 10pt]{IEEEtran}
\usepackage{amsfonts, latexsym, amsmath, amssymb}
\usepackage{bbm}
\usepackage[latin1]{inputenc}
\usepackage[english]{babel}
\usepackage{graphicx}
\usepackage{url}
\usepackage{subfigure}
\usepackage[only,llbracket, rrbracket]{stmaryrd}
\usepackage{psfrag} 
\usepackage{multirow}

\newtheorem{lemma}{Lemma}
\newtheorem{thm}{Theorem}
\newtheorem{rem}{Remark}

\newcommand{\discretAB}[2]{\left \{ #1, \ldots, #2 \right \}}

\title{On the Threshold of Maximum-Distance Separable Codes}
\author{
\IEEEauthorblockN{Bruno Kindarji\IEEEauthorrefmark{1}\IEEEauthorrefmark{2}}
\IEEEauthorblockA{
\IEEEauthorrefmark{1}
Sagem S\'{e}curit\'{e}\\
Osny, France}
\and 
\IEEEauthorblockN{G\'{e}rard Cohen\IEEEauthorrefmark{2}}
\and
\IEEEauthorblockN{Herv\'{e} Chabanne\IEEEauthorrefmark{1}\IEEEauthorrefmark{2}}
\IEEEauthorblockA{
\IEEEauthorrefmark{2}Institut T\'{e}l\'{e}com \\
T\'{e}l\'{e}com ParisTech \\
Paris, France}
}

\begin{document}
\date{}
\maketitle
\pagestyle{empty}
\begin{abstract}
Starting from a practical use of Reed-Solomon codes in a cryptographic scheme published in Indocrypt'09, this paper deals with the threshold of linear $q$-ary error-correcting codes. The security of this scheme is based on the intractability of polynomial reconstruction when there is too much noise in the vector. Our approach switches from this paradigm to an Information Theoretical point of view: is there a class of elements that are so far away from the code that the list size is always superpolynomial? Or, dually speaking, is Maximum-Likelihood decoding almost surely impossible?

We relate this issue to the decoding threshold of a code, and show that when the minimal distance of the code is high enough, the threshold effect is very sharp. In a second part, we explicit lower-bounds on the threshold of Maximum-Distance Separable codes such as Reed-Solomon codes, and compute the threshold for the toy example that motivates this study.
\end{abstract}
\section{Introduction}
In \cite{BringerCCK09}, Bringer~\textit{et al.} proposed a low-cost mutual authentication protocol, that uses a Reed-Solomon code structure. This protocol is pretty simple: Bob owns two secret polynomials $P_b, P_b'$ of degree less than $k$ known only by Alice; to authenticate herself to Bob, Alice proves the knowledge of $P_b$ by sending $\left\langle i,P_b(\alpha_i)\right\rangle$ where $\alpha_i$ is the $i$-th element of a $\mathbb{F}_q$. Bob proves his identity by replying with $\left\langle P_b'(\alpha_i)\right\rangle$. This protocol is made such that if Alice speaks to a lot of persons, it is hard to trace Bob out of all the conversations, and it is hard to impersonate Alice (or Bob).
The security of the protocol is based on an algorithmic assumption, saying that the polynomial reconstruction problem is hard for the vectors of $\mathbb{F}_q^n$ that are far enough from the code. Indeed, the best known algorithms solving polynomial reconstruction are those of Guruswami-Sudan \cite{Guruswami02reflectionson} and, on a related problem, Guruswami-Rudra \cite{GuruRudra09}, which can basically reconstruct a polynomial given $\sqrt{kn}$ correct values.

This algorithmic security result is somehow unsatisfying, for it is possible to exhibit better decoding algorithm. We therefore take interest in the information-theoretic aspect of such a problem.

The solution of the problem raised by \cite{BringerCCK09} is to look at the output of a list-decoder centered around the received values, and to output the possible polynomials as candidate values for $P_b$ or $P_b'$. Our approach consists in looking at a usually ignored side of list-decoding, which is to look at the radii $r$ such that list-decoding a word with radius $r$ provides a list that is always lower-bounded by a large enough number. This differs from the literature concerning list-decoding, which usually looks for radii for which the size is always upper-bounded by a maximum list size, or tries to exhibit a counter-example.

The ``large enough'' list size can be obtained easily by imposing that Maximum-Likelihood Decoding to be most improbable. For that, we focus on the all-or-nothing behaviour of the ML decoder. Inspired by percolation theory \cite{Grimmett97percolation}, and code-applied graph theory \cite{TilZem00}, we will show how it is possible to conservatively estimate, before, after, and around a threshold, the all-or-nothing probability of ML decoding.

\section{The Threshold of a Code}
The existence of a threshold is motivated by the classical question of percolation : given a graph, with a source, and a sink, and given the probability $p$ for a ``wet'' node of the graph to ``wet'' an adjacent node, \textit{what is the probability for the source to wet the sink}? It appears that this probability has a threshold effect; in other words, there exists a limit probability $p_c$ such that, if $p>p_c$, then the sink is almost surely wet, and if $p<p_c$, then the sink is almost never wet. The threshold effect is illustrated in Fig~\ref{fig:thres}.

This question can be transposed into the probability of error-correcting a code. Given a proportion of errors $p$, with a decoding algorithm, what is the probability of correctly recovering the sent codeword? It was shown in \cite{Zemor93} that for every binary code, and every decoding algorithm, this probability also follows a threshold.

In this paper, we show that this property also applies to $q$-ary codes. In the following part, we show that the threshold behaviour that was seen on binary codes can be obtained again.

\subsection{The Margulis-Russo Identity}

The technique used to derive threshold effects in discrete spaces is to integrate an isoperimetric inequality; for that, the Margulis-Russo identity is required.

Let $H = \{0,1\}^n$ be the Hamming space; the Hamming distance $d(x,y)$ provides the number of different coordinates between vectors $x$ and $y$. Consider the measure $\mu_p : H \rightarrow [0,1]$ defined by $\mu_p(x) = p^{w(x)}(1-p)^{n-w(x)}$ where $w(x)$ is the Hamming weight of $x$. The number of limit-vectors of a subset $A \subset H$ is a function defined as $h_A(x) = |B(x,1) \cap \overline A|$ for $x\in A$.

For $A\subset H$ such that $A$ is increasing (\textit{i.e.} if $x\in A$, and $y \geq x$, then $y\in A$ with $\geq$ defined component-wise), Margulis and Russo showed :

$$ \frac{d \mu_p (A)}{dp} = \frac{1}{p} \int_A h_A(x) d\mu_p(x)$$

Let $q \in \mathbb{N}, q>2$. This section shows that this equality is also true in $H_q=\{0, ... q-1\}^n$.

For a vector $x\in H_q$, the support of $x$ is the set of all its non-null coordinates, \textit{i.e.} $supp(x) = \{i\in \discretAB{1}{n} : x_i \neq 0\}$. Define the measure function $\mu_p(x) = \left ( \frac{p}{q-1} \right )^{w(x)} (1-p)^{n-w(x)}$ with $w(x) = |supp(x)|$ the weight of $x$. This definition is consistent with a measure, as $\mu_p(H_n) = \sum_{x\in H_q} \mu_p(x) = 1$.

Note the inclusion $\subset$ to be the relation between a set and a (general) subset (\textit{i.e.} for all $X$, $X\subset X$). The support inclusion generalises the component-wise $\leq$ that was used in the binary case.

\begin{lemma}[Margulis-Russo Identity over $q$-ary alphabets] Let $A$ be an increasing subset of $H_q$, \textit{i.e.} such that if $y\in A$, for all $x \in H_q$ such that $supp(y) \subset supp(x)$, then $x \in A$. Then

$$ \frac{d \mu_p (A)}{dp} = \frac{1}{p} \int_A h_A(x) d\mu_p(x)$$

\end{lemma}

\IEEEproof
The proof of this lemma is an adaptation of Margulis' proof in \cite{margulis74}. For this, we use the notation:
\begin{itemize}
\item $[A,B] = |\{x,y\}\in A \times B : d(x,y)=1|$ where $A,B\subset H_q$, is the number of links from $A$ to $B$
\item for $k\in \discretAB{0}{n}$, $Z_k = \{x \in H_q : w(x)=k \}$,
\item for $A\subset H_q$, $A_k = A \cap Z_k$ ($A$ is the reunion of the $A_k$);
\item $D_k = \sum_{x\in A_k} h_A(x)$ is the number of limit-vectors next to elements of weight $k$.
\end{itemize}

Trivially, $D_k = [A_k, Z_{k+1} - A_{k+1}] + [A_k, Z_{k-1} - A_{k-1}] + [A_k, Z_k - A_k]$. We now note that :

\begin{itemize}
\item $[A_k, Z_{k-1}] = |A_k| k$, as to go from $A_k$ to $Z_{k-1}$, the only way (in one move) is to put one coordinate to $0$;
\item $[A_k, Z_{k+1}] = |A_k| (n-k)(q-1)$ with the same reasoning;
\item $[A_k, Z_k - A_k ] = [A_k, Z_{k+1} - A_{k+1}] = 0$ as $A$ is increasing.
\item Combining these equalities, we get $[A_k, A_{k+1}] = |A_k| (n-k)(q-1)$;
\item $[A_k, Z_k] = 0 $ as it is necessary to put a non-null coordinate to $0$ and a null one to $\{1, ... q-1\}$.
\end{itemize}

Finally $D_k = [A_k, Z_{k-1}] - [A_k, A_{k-1}] = k |A_k| - (n-k+1)(q-1)|A_{k-1}|$ for $k>0$ and $D_0 = 0$ (or $A=H_q$).

Back to the identity desired, we observe that 
\begin{eqnarray*}
\int_A h_A(x) d\mu_p(x) & = & \sum_{k=0}^{n} \sum_{x\in A_k} h_A(x) (\frac{p}{q-1})^k (1-p)^{n-k} \\
								& = & \sum_{k=0}^{n} D_k \left (\frac{p}{q-1}\right)^k (1-p)^{n-k}
\end{eqnarray*}
$$
\begin{array}{l}
\displaystyle								 =  \sum_{k=1}^{n} \left (k |A_k| - (n-k+1)(q-1)|A_{k-1}| \right )  \\
\hspace{4cm} \displaystyle  \cdot \left (\frac{p}{q-1}\right)^k (1-p)^{n-k} \\
								 =  \sum_{k=0}^{n} |A_k| (k - p \frac{n-k}{1-p}) \left (\frac{p}{q-1}\right)^k (1-p)^{n-k} \\
\text{on the other hand,}\\
\frac{d \mu_p (A)}{dp} 	 =  \sum_{k=0}^{n} |A_k| \frac{d}{dp} \left ( \left (\frac{p}{q-1}\right)^k (1-p)^{n-k} \right ) \\
								 =  \sum_{k=0}^{n} |A_k| \left(\frac{p}{q-1}\right)^k (1-p)^{n-k} \left (\frac{k}{p} + \frac{-(n-k)}{1-p} \right)  
\end{array}
$$
Hence the identity.
\endproof

This lemma shows that the Margulis-Russo identity is also true on $\{0 ... (q-1)\}^n$; it was the keystone of the reasoning done in \cite{TilZem00} to show an explicit form of the threshold behaviour of Maximum-Likelihood Error Correction.

\subsection{A Threshold for Error-Decoding $q$-ary codes}

In the following, we use $\varphi(t) = \frac{1}{\sqrt{2\pi}} e^{-\frac{t^2}{2}}$ the normal distribution, $\Phi(x)  = \int_{-\infty}^{x} \varphi(t) dt$ the accumulate normal function, and $\Psi(x) = \varphi(\Phi^{-1}(x))$ (so that $\forall x, \Psi(x) \cdot \Phi'^{-1}(x)=1$).

A monotone property is a set $A\subset H_q$ such that $A$ is increasing, or $\overline{A}$ is increasing.

\begin{thm}
	Let $A$ be a monotone property of $H_q$. Suppose that $\forall x\in A, h_A(x)=0$ or $h_A(x) \geq \Delta$.
	
	Let $\theta \in [0,1]$ be (the unique real) such that $\mu_\theta(A) = \frac{1}{2}$. Let $g_\theta(p) = \Phi \left ( \sqrt{2\Delta} (\sqrt{-\ln \theta} - \sqrt{-\ln p})\right )$.
	
	Then the measure of $A$, $\mu_p(A)$ is bounded by :
	$$\begin{array}{rcl c}
		\mu_p(A) & \leq &  g_\theta(p) & \text{ for } p\in ]0;\theta] \\
		\mu_p(A) & \geq &  g_\theta(p) & \text{ for } p\in [\theta;1[ 
	\end{array} $$ 
\end{thm}

\textit{Sketch of Proof}\\
The proof is exactly the same as the one from \cite{TilZem00}. The whole idea is to derive the upper-range:
$$\int_A \sqrt{h_A} d\mu_p \geq \sqrt{2 \ln \frac{1}{p}} \Psi(\mu_p(A))$$
The integration of this equation, together with the Margulis-Russo lemma, gives the result.
\endproof

To conclude this part, we remark that the non-decoding region of a given point, for a $q$-ary code, is an increasing region of $\mathbb{F}_q^n$. For linear codes, this non-decoding region can always be translated to that of $0$ without loss of generality; let $A_0 = \{x \in \mathbb{F}_q^n \text{ s.t. } \exists c \in C, c \neq 0 : d(x,c)\leq d(x,0)\}$. The probability of error decoding of $C$ is then $\mu_p(A_0)$.

For $x\in \mu_p(A_0)$, we show that either $h_{A_0}(x)=0$, or $h_{A_0}(x)\geq \frac{d}{2}$. Indeed, if $h_{A_0}(x)>0$, then $x$ is nearer to a non-null codeword $c$ than to $0$. Then all the vectors obtained by replacing one of the coordinates of $x$ by $0$ are out of $A_0$; in particular, $h_{A_0}(x) \geq d(x,0)$. Let $d_c=d(c,0)$ be the weight of $c$; as $x$ is nearer to $c$ than to $0$, $d(x,0) \geq \frac{d_c}{2}$. Thus the previous assertion.

Combining the previous results, we just showed that for any $q$-ary code, the probability of error is, as for binary codes, bounded by a threshold function. This can be expressed by the following theorem, which has the same form as the one showed in \cite{TilZem00}:
\begin{thm}
	Let $C$ be a code of any length, and of minimal distance $d$. Over the $q$-ary symmetric channel, with transition probability $p$, the probability of decoding error $P_e(p)$ associated with $C$ is such that there exists a unique $p_c \in ]0;1|$ such that $P_e(p_c) = \frac{1}{2}$, and $P_e$ is bounded by:
$$	 	P_e(p) \lesseqqgtr  1 - \Phi ( \sqrt{d} (\sqrt{-\ln(1-p_c)} - \sqrt{-\ln(1-p)})  )$$
	The upper-bound ($\leq$) is true when $p\in ]0;p_c]$; the lower-bound ($\geq$) is true when $p\in [p_c;1[$.
\end{thm}

Even though linearity was used not to lose any generality previously, it is not a requirement for this theorem. Indeed, the bounding equations are true for every codeword $c$ by replacing $d$ by $\min_{c'\in C, c`\neq c} d(c,c')$. Assuming that the codewords sent are distributed in a uniform way over $C$, we thus obtain this result.

\begin{figure}
\includegraphics[width=\columnwidth]{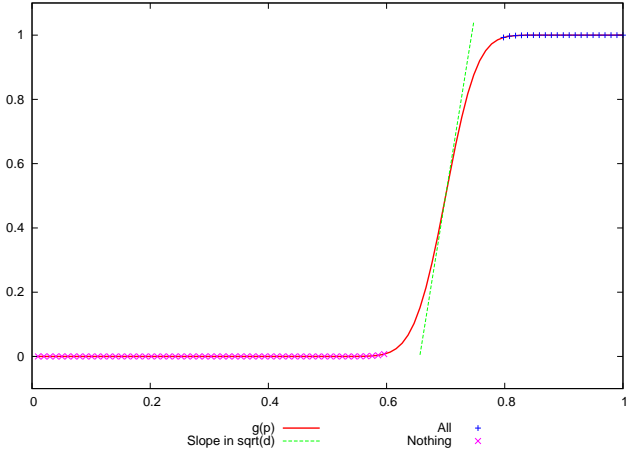}
\caption{Illustration of the threshold effect, $d=400$, $p_c=0.7$}
\label{fig:thres}
\end{figure}

The behaviour of this function is illustrated in Fig~\ref{fig:thres}. Around $p \approx 0$ (actually, for all $p<p_c - \epsilon$...), $P_e$ is extremely flat around 0; around $p \approx 1$ (and, symmetrically, for all $p > p_c + \epsilon$, $P_e$ is extremely flat around 1. Finally, around the threshold $p_c$, the slope is $\frac{\sqrt{d}}{\sqrt{2\pi}(1-p_c)}$, which is almost vertical when the minimal distance $d$ is large.

\section{Explicit Computation of the Threshold for Maximum-Distance Separable Codes}

In this section, we only take interest in linear codes over $\mathbb{F}_q^n$.

\subsection{Another Estimation of the Decoding Threshold}
\label{sec:esti}

By linearity, we can again without loss of generality assume that the sent codeword was the all null vector. 
It is possible to have a rough estimation of the probability of wrongly decoding with crossover probability $p$ correctly a vector by computing the proportion of vectors $x\in \mathbb{F}_q^n$ of weight less or equal to $np$ that are closer to a non-null codeword than to $0$. Let $g(p)$ be this proportion.

$$g(p) = \frac{|\left \{ x : \text{ s.t. } \exists c\in C, c\neq 0 : d(x,c)\neq w(x) \leq np  \right \}|}{|\left \{ x : w(x) \leq np \right \}|}.$$

Let $vol(q,n,t) = \frac{1}{n}\log_q \left ( |B(t)|\right)$, where $B(t)$ is the Hamming ball of radius $t$, ( for example, centered on $0$) in $\mathbb{F}_q^n$. It is well known that when $t\leq \frac{q-1}{q}$, $vol(q,n,t) = H_q(\frac{t}{n}) + o_n(1)$, where $H_q(x)=-x \log_q x - (1-x) \log_q (1-x) + x \log_q (q-1)$ is the $q$-ary entropy of $x\in [0,1]$.

To compute the numerator, we suggest, for each codeword $c\in C$ that has a weight between $d$ and $2pn$, to compute the number of vectors $x$ that are nearer to $c$ than to $0$. This number actually only depends on the weight of $c$, and will be noted $\nu_{pn}(w(c))$. As there are $A_{w(c)}$ codewords of weight $w(c)$ in the code (with the standard notation), the function $g(p)$ can be approximated by:
\begin{equation}
g(p)\leq \frac{\sum_{l=d}^{2pn} A_l \nu_{pn}(l) }{q^{n vol(q,n,pn)}}
\label{eq:approx}
\end{equation}

\begin{figure}[hbt]
\begin{center}
	\includegraphics[width=0.7\columnwidth]{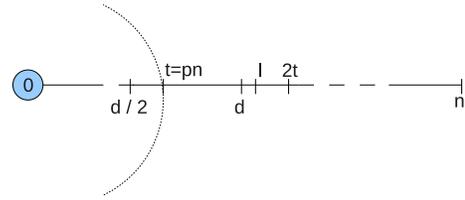}
\end{center}
\caption{Different quantities used in Eq~\ref{eq:approx}}
\label{fig:repart}
\end{figure}

The different quantities used in this equation are illustrated in Fig~\ref{fig:repart}.

$\nu_t(w)$ is explicited hereafter. Let $c$ be a codeword of weight $w$. Let $x\in \mathbb{F}_q^n$ be a vector with the following constraints:
\begin{itemize}
	\item $d(x,0)\leq t$, \textit{i.e.} $x$ is the result of the transmission of $0$ with at most $t$ errors.
	\item $d(x,0) \geq d(x,c)$, \textit{i.e.} $x$ is wrongly decoded.
\end{itemize}

We note $\alpha$ the number of coordinates $i$ in $x$ such that $x_i\neq c_i$ and $x_i=0$; $\beta$ is the number of coordinates $i$ such that $x_i \neq c_i$ and $x_i \neq 0$; $\gamma$ is the number of coordinates $i$ such that $x_i \neq c_i$ and $c_i = 0$. 

The previous constraints on $x$ can be rewritten into the system $(S)$:
$$(S): \left \{ \begin{array}{rl}
	1) & 0 \leq \alpha, \beta \leq w \\
	2) & 0 \leq \gamma \leq n - w \\
	3) & \gamma \leq t + \alpha - w \\
	4) & \beta + \gamma \leq t \\
	5) & 2 \alpha + \beta \leq w 
\end{array} \right .$$

We then obtain $$\nu_t(w) = \sum_{\alpha, \beta, \gamma} {w \choose \alpha + \beta}{\alpha + \beta \choose \beta}(q-2)^\beta{n - w \choose \gamma} (q-1)^\gamma.$$

\begin{rem}
It is easy to see that $\nu_t(w)$ is at most the volume of a ball of radius $w-\frac{d}{2}$; this estimation will be used in the next part.
\end{rem}

\subsection{Application to MDS codes}

Maximum-Distance Separable (\textbf{MDS}) Codes are codes such that their dimension $k$ and minimal distance $d$ fulfil the Singleton bound, so that:
$$ k + d = n - 1. $$

A well known family of MDS codes are the Reed-Solomon codes, for which a codeword is made of the evaluation of a degree $k-1$ polynomial over $n$ field elements $\alpha_1, \ldots, \alpha_n$. Reed-Solomon codes over $\mathbb{F}_q$ can have a length up to $q-1$, but shorter such codes are also MDS.

For MDS codes, the number $A_l$ of codewords of given weight is known. This number is:

$$A_{n-i} = \sum_{j=1}^{n-1} (-1)^{j-i} {n \choose j }{j \choose i} (q^{k-j}-1)$$

From this identity, it is easy to derive the more usable formula:
\begin{equation}
A_l = {n \choose l } \sum_{j=0}^{l-d} (-1)^j {l \choose j}(q^{1+l-d-j}-1)
	\label{eq:enum}
\end{equation}

It is now possible to approximate quite nicely the error probability while under the threshold - indeed, the numerator and denominator are correct as long as a vector $x$ is not close to 2 different codewords with a weight in the range $[d;pn]$, \textit{i.e.} as long as the list of codewords at a distance less than $pn$ from $x$ is reduced to a single element.

\subsection{Short MDS Codes over Large Fields}

We now focus on the specific problem presented in the Introduction, and motivated by the beckoning and authentication protocol from \cite{BringerCCK09}. This setting is characterized by the following:
\begin{itemize}
	\item The underlying code is a Reed-Solomon over a field $\mathbb{F}_q$;
	\item The field size $q$ is very large for cryptographic reasons;
	\item The code length $n$ is very short (with respect to $q$) as $nq$ is the size of embedded low-cost devices' memory.
\end{itemize}

This application fits into the framework depicted in the previous sections. Moreover, the information ``$n$ much smaller than $q$'' ($n=o(q)$) enables to compute an asymptotic first order estimation of the threshold in such codes.

Indeed, if $g(p)\leq f(p)$, then $g^{-1}(\frac{1}{2}) \geq f^{-1}(\frac{1}{2})$. We now compute an upper-bound on $g(p)$, to derive an estimation on the threshold $\theta$. More precisely, we aim at computing $\iota(p)$ the first-order value of $\log_q \left(g(p)\right)$; then, $\iota^{-1}(0)$ is a lower-approximation of the threshold.

To estimate the weight enumerator $A_l$, we use formula ($\ref{eq:enum}$) to derive $$A_l \leq (l-d){n \choose l} 2^l q^{1+l-d} \leq n 2^{n + l} q^{1+l-d}.$$

The number of targetted vectors for each codeword $\nu_t(l)$ is not easy to evaluate; we note its first order development $\log_q \nu_t(l) = n \mu(l,t) + o_q(1)$, so that $\nu_t(l) \leq q^{n \mu(l,t)} \cdot o_q(q)$. (Here, the term $o(q)$ is a bounded by a polynomial in $n$.) We know that \begin{equation}
0 \leq n \mu(l,t) \leq l-\frac{d}{2}
	\label{eq:enc_mu}
\end{equation}

Combining these elements with equation (\ref{eq:approx}), we obtain $g(p) \leq \sum_{l=d}^{2pn}o(q) q^{1+l-d + \mu(l,pn) - n vol(q,n,pn)}$.

As $vol(q,n,t) = H_q(\frac{t}{n}) + o_n(1) = \frac{t}{n} + o_q(1)$, the first order of $g(p)$ is bounded by: $\log_q g(p) \leq \max_{l\in [d,pn]} \left (1 + l-d - pn + n \mu(l,pn) \right )+ o_q(1)$.

The bounding (\ref{eq:enc_mu}) of $\mu$ shows that the right-hand side of this inequality is between $1 + pn - d$ and $1 + 3 pn - \frac{3d}{2}$, which shows that the threshold $g^{-1}(\frac{1}{2})$ is asymptotically between $\frac{\delta}{2}$ and $\delta$.

Unfortunately, a more precise evaluation of $\mu$ strongly depends on the context. Indeed, according to Section \ref{sec:esti},
$$\nu(l,t) = o_q(q) \cdot \max_{\alpha, \beta, \gamma : (S)} q^{\beta + \gamma} {n - l \choose \gamma}{l \choose \alpha + \beta}{\alpha + \beta \choose \beta}.$$
This maximum can be obtained by evaluating the term to be maximized on all vertices of the polytope defined by the system $(S)$ ($(S)$ is made of 9 inequalities of 3 unknown, the vertices are obtained by selecting 3 of these equations, thus at most ${9 \choose 3} = 84$ vertices); however, it is not possible to exhibit here a general answer as the solution depends on the minimal distance of the code, \textit{i.e.} on the rate of the Reed-Solomon code.

\subsection{Numerical Application to a $(2048,256,1793)_{2^{64}}$ MDS Code}

In the case of a code over a finite field of reasonable dimension, it is possible to exactly compute the ratio that approximates the Maximum Likelihood threshold. However, the exact threshold cannot be easily computed yet; it is still an open problem related to the list-decoding capacity of Reed-Solomon codes.

We therefore used the NTL open-source library \cite{ntl} to compute the values $A_l$, $\nu_t(l)$ and $|B(t)|$ in order to have an accurate enough approximation of the the function $g(p)$ described earlier. 
The parameters are those that were proposed in \cite{BringerCCK09}, and show that the decoding threshold of such a code is between $0.8$ and $0.875$.

The slope around the threshold is around 115, so for $p$ ``small'' (in fact, a bit smaller than $p_c$) $g(p)$ is very near to $0$, while as $p$ goes to $1$, $g(p)$ is much greater than the maximum probability of $1$. This was predicted earlier, and expresses the fact that the list-size of radius $pn$ is always greater than 1. The threshold value $g^{-1}(\frac{1}{2}) \approx \iota^{-1}(0)$ is a lower-bound for the threshold of the code, though the intuition says that this lower-bound is pretty near to the real threshold.

\section{Conclusion}

As a conclusion, let us look back to the starting point of our reasoning. The initial goal was to revise the conditions of security of the construction depicted in \cite{BringerCCK09}: from a received vector $x$ of $\mathbb{F}_q^{n}$, for what parameters is the size of the list of radius $pn$ exponentially large? This problem can be reduced to that of the threshold probability of a linear error-correcting code. Indeed, below the threshold of the code, when the minimal distance of the code is large enough, the error decoding probability of the code is exponentially small, and it is exponentially close to 1 above the threshold. For our class of parameters, ensuring that the error rate is above the threshold is enough to show the security of the scheme.

We then showed that the threshold behaviour can be explicited for $q$-ary codes as well as for binary codes; we then explicited a lower-bound on the threshold of MDS codes.

Applying these results to the initial problem, we show that the threshold for a (highly) truncated Reed-Solomon code over a finite field $\mathbb{F}_{2^{64}}$ is very near to normalized the minimal distance $d=n-k+1$ of this code. As a conclusion, to switch from an algorithmic assumption (the hardness of the Polynomial Reconstruction Problem, see \cite{DBLP:conf/icalp/KiayiasY02}) to Information-Theoretical security, we recommend to raise the dimension $k$ of the underlying code. This lowers the decoding threshold of the code; the downside is that storage of a codeword is more costly. 

\section{Acknowledgements}

We thank Gilles Zémor for the useful comments and fruitful discussions.

\bibliographystyle{IEEEtran}
\bibliography{codes,communications,indo09}
\end{document}